\documentclass[%
 reprint,
 amsmath,amssymb,
 aps,
]{revtex4-1}

\usepackage{graphicx}
\usepackage{dcolumn}
\usepackage{bm}

\begin{document}


\title{Complex ordering in spin networks:
Critical role of adaptation rate for dynamically evolving
interactions}

\author{Anand Pathak}
\email{anandp@imsc.res.in}
\author{Sitabhra Sinha}%
 \email{sitabhra@imsc.res.in}
\affiliation{%
 The Institute of Mathematical Sciences,
 CIT Campus, Taramani, Chennai 600113, India.
 }%

%
%

\date{\today}

\begin{abstract}
Many complex systems can be represented as networks of dynamical
elements whose states evolve in response to interactions with
neighboring elements, noise and external stimuli. The collective
behavior of such systems can exhibit remarkable ordering phenomena
such as {\em chimera order} corresponding to coexistence of ordered
and disordered regions. Often, the interactions in such systems can
also evolve over time responding to changes in the dynamical states of
the elements. Link adaptation inspired by Hebbian learning, the
dominant paradigm for neuronal plasticity, has been earlier shown to
result in structural balance by removing any initial
frustration in a system that arises through conflicting interactions.
Here we show that the rate of the adaptive dynamics for the
interactions is crucial in deciding  the emergence of different
ordering behavior (including chimera) and frustration in networks of
Ising spins.
In particular, we observe that small changes in the link adaptation rate 
about a critical value
result in the system exhibiting radically different energy landscapes,
viz., smooth landscape corresponding to balanced systems seen for
fast learning, 
and rugged landscapes corresponding to frustrated systems seen for slow
learning.
\end{abstract}

\pacs{Valid PACS appear here}
\maketitle

\section{\label{sec:Introduction}Introduction}

Many natural and technological complex systems can be described
as networks connecting a large number of elements whose states
evolve in time~\cite{Newman2010,Barrat2008}. The collective behavior
of a system resulting from interactions between its components
can exhibit non-trivial features, including critical
phenomena~\cite{Dorogovtsev2008}. For instance, a system of binary-state 
elements (e.g., representing individuals having opposing opinions on an issue) 
connected through a network having modular organization can
exhibit ordering dynamics at very distinct time-scales~\cite{Pan2009} and
under certain circumstances, self-organize into locally aligned clusters that
correspond to the communities of the network (i.e., subnetworks
characterized by a significantly higher connection density
compared to the overall density of the network)~\cite{Dasgupta2009}.
In many situations, the links of the network (representing the
interactions) can also
evolve over time as a result of the changes in the states of the
components that they connect. Such connections may not only be
characterized by weights (indicating the strength of interaction) but
also sign (representing the nature of the interaction). For instance,
in the context of a network of synaptically-connected model neurons, 
positive links may correspond to
excitatory interactions (whereby activation of one element can result
in activation of other connected elements) while negative links
can give rise to inhibition (i.e., activation of an element tends to suppress
subsequent activation of neighboring elements)~\cite{Hertz1991}.

The occurrence of negative links can lead to the emergence of {\em
frustration} 
because of the existence of inconsistent relations within cycles in the
network~\cite{Mezard1987,Fischer1991}. A network is said to be
structurally balanced if its positive and negative
links are arranged such that frustration is absent. Such a network
can always be represented as two communities, with interactions within
each community being exclusively positive while those between
communities can only be negative~\cite{Cartwright1956}.
This is of interest not only for social systems in the context of
which the concept of balance was first introduced~\cite{Heider1946},
but also for biological systems. For instance, it has been recently
observed that the resting human brain is organized into
a pair of dynamically anti-correlated subnetworks~\cite{Fox2005}. This
suggests that the network responsible for this behavior may be
structurally balanced. This possibility is intriguing in view
of the observation of non-trivial collective behavior in 
balanced networks of binary-state dynamical elements, such as, the coexistence 
of
ordered and disordered regions referred to as ``chimera''
order, in the presence of an external field~\cite{Singh2011}. 
The process by which networks can evolve to a
balanced configuration has been explored by considering a link
adaptation process inspired by Hebb's principle, the basis behind
neural plasticity~\cite{Hebb49}. According to this mechanism, weights
associated with each link change in proportion to
the correlation in the activity of the connected elements.
Therefore, any initial frustration in the network can be removed 
by modifying
the connection weights in accordance
with the dynamical states of the elements. Systems undergoing such
link adaptation in the presence of environmental noise show high
variability in the convergence time required to reach the 
balanced state~\cite{Singh2014}. 

In this paper we consider how the rate of dynamical evolution of interactions
in accordance with the Hebbian adaptation principle affects the
collective behavior of the network.  
Starting from a fully connected network that is structurally balanced
and introducing noise and external field so that the system exhibits
chimera order, we show that different rates
of adaptation can result in distinct outcomes. 
Fast learning rates result in persistence of the chimera regime, 
although the balanced network now comprises communities with
asymmetric sizes, viz., the ordered sub-network containing a much
larger fraction of elements of the network.
Slow learning, on the other hand, results in a completely ordered state
with the interactions becoming only of positive nature.
On removing the external field, fast learning results in retrieving a network
that is similar to the original one, i.e., 
structurally balanced with communities of almost equal size. However,
slow learning gives rise to a fully frustrated network exhibiting disorder. 
We observe that there exists a critical interval for values of the
learning rate, around which small changes result in the system
converging to distinct final states. In the following sections we
first discuss the model and adaptation dynamics, followed by
description of the results of numerical simulations. We conclude with
a brief summary of our results and a discussion of their implications.

\section{The Model}
For the purpose of investigating how collective behavior of a complex
system is affected
by adaptive dynamics of interactions, we use one of the well-known spin
models of statistical physics which are generic systems for
analyzing cooperative phenomena. In particular, we use the
binary-state Ising spin, the spin orientations (``up'' or ``down'') 
representing a pair of mutually exclusive choices. The simplicity of
the model makes it applicable to not just magnetic materials (in the
context of which it was originally proposed) but any system where the
elements choose between the two competing states based on interactions
with neighboring elements, noise (represented by thermal fluctuations
characterized by a temperature $T$) and external stimulus (often
represented by a magnetic field $H$).
The interactions $J_{ij}$ between any pair of spins ($i$, $j$) can be
either positive (promoting connected spins to have the same state)
or negative (promoting connected spins to have opposite states)
in nature. For instance, in neuronal networks, one can view the
neurons as binary-state devices that are either firing (active) or quiescent
(inactive). Correspondingly, as pointed out earlier, the
excitatory and inhibitory connections between neurons can be
represented by positive and negative interactions, respectively.
At a different scale, one can view genetic regulatory networks in a
similar vein, with genes being in either of two states, viz., 
getting expressed (active) or not(inactive). In this setting, genes
promoting the expression of other genes correspond to positive
interaction, while inhibiting the expression of other genes correspond
to negative interactions.
In a different context, the interactions between Ising spins can also
be used to represent social intercourse~\cite{facchetti}. Here, the
spin states are considered to be analogous to two competing opinions,
while the interactions may represent the nature of the relation between 
a pair of individuals - positive corresponding to affiliative and
negative corresponding to antagonistic relations.

We consider a system of $2 N$ globally coupled Ising spins
arranged into two sub-populations (called modules or communities)
each having $N$ spins, at a constant temperature $T$. For the
simulations reported below we have chosen $N = 100$.
The
interaction between a pair of spins belonging to the same module 
is positive having strength $J$ ($>0$), while that between spins belonging
to different modules is negative with strength
$-J^{\prime}$ (where $J^{\prime} >0$). In the absence of
an external field and thermal fluctuations (i.e., $H = 0$, $T = 0$), the two
modules will be completely ordered in opposite orientations.
When subjected to an external field of strength $H$, the energy
of the system is described by
\begin{equation}
E = -J\sum_{{i,j,\alpha}}\sigma_{i\alpha}\sigma_{j\alpha}
-J'\sum_{i,j,\alpha,\beta} \sigma_{i\alpha} \sigma_{j\beta}
-H\sum_{i,\alpha} \sigma_{i\alpha},
\label{eq1}
\end{equation}
where $\sigma_{i\alpha} = \pm 1$ is the Ising spin on the $i$-th node
($i, j = 1, 2, \ldots N$) in the $\alpha$-th module ($\alpha, \beta
=1, 2$). 
As every spin interacts with every other spin, a mean-field
treatment should describe the system accurately. For convenience we
define $a=J(N-1)/2$ and $b= J^{\prime} N$ as system parameters. The
state of all spins in the system are updated stochastically
at discrete time-steps using the Metropolis Monte
Carlo (MC) algorithm with temperature $T$ expressed in units
of $k_B T/ J$ where $k_B$ is the Boltzmann constant.
The initial state of the system we have chosen is one exhibiting chimera
order~\cite{Singh2011}, to obtain which we use
$a=1$, $b=H=10$ and $k_{B}T/J=5$. 

We also allow the possibility that the interaction strengths can
change over time through an adaptation process inspired by the
principle of Hebbian learning, a classical concept in the area of
neural networks~\cite{Hebb49}. Colloquially often described as
``neurons that fire together, wire together", in this mechanism the
strength of synaptic connection between a pair of neurons is
incremented when the two exhibit correlated activation.
In Ref.~\cite{Singh2014}, this idea was used to propose a link adaptation rule 
that applies to a system of spins coupled via positive, as well
as, negative interactions as:
\begin{equation}
J_{ij}(t + 1) = (1 - \epsilon)J_{ij}(t) + \epsilon \sigma_i(t)\sigma_j(t),
\end{equation}
which is applied after every MC step. The adaptation rate, $\epsilon$,
decides the time-scale over which the interaction strength changes
relative to the spin dynamics.
Starting with an initially frustrated spin system, implementing the
above $J_{ij}$ dynamics in absence of thermal fluctuations (i.e.,
$T=0$) results in the system converging to a structurally balanced
state~\cite{Singh2014}. This can be intuitively understood in terms of 
changes in the energy landscape over which the state of
the spin system evolves. Frustrated systems have rugged energy
landscapes comprising a large number of local minima (a fact which is
exploited in neural network models of associative memory where these
minima are used to embed desired patterns one wishes the network to
``memorize"~\cite{Hopfield82,Pradhan11}. Thus,
beginning at any randomly chosen initial state of spin configurations, 
the spin dynamics drives the system into the nearest local energy
minimum. The subsequent evolution of the $J_{ij}$s
converts this into the global minimum of the system.
However, in the presence of noise ($T >0$),
fluctuations in the spin dynamics can prevent the system
from being trapped in any state for sufficiently
long duration. Thus, the adaptation dynamics fails to alter the
energy landscape sufficiently so as to turn the configuration
into the global minimum. Therefore, with increasing temperature, an extremely 
long time may be required to reach structural balance.

In this paper we use a modified form of the Hebb-inspired adaptation
rule for interaction strengths that was introduced in
Ref.~\cite{Singh2014}. This is to take into account the asymmetry in the
strengths of positive and negative interactions necessary for
observing chimera order (typically $J^\prime \simeq 10 J$)~\cite{Singh2011}.
We change the adaptation rule such that the upper and lower
bounds for the evolving interactions have the same values as that of
the positive and negative interactions (respectively) which give 
rise to the chimera ordered state.
Thus after each MC step, interaction strengths are changed according
to:
\begin{equation}
 J_{ij}(t+1) =
 J_{ij}(1-\epsilon)+\epsilon[(J+J^{\prime})\frac{\sigma_i\sigma_j}{2}+\frac{1}{2}(J-J^{\prime})].
\label{hebbrulenew}
\end{equation}

The state of the system at any time is characterized by two
quantities. One of these is the
frustration associated with the interactions, measured by calculating the
fraction of frustrated triads (i.e., connected sets of 3 spins with an
odd number of negative interactions) in the system:
\begin{equation}
F = \sum_{i\neq j\neq k}\frac{ J_{ij}J_{jk}J_{ki}}{^NC_3}, \;\;
\forall J_{ij}J_{jk}J_{ki}<0,
\end{equation}
which varies between $0$ (no frustration) and $1$ (completely
frustrated).

\begin{figure}[tbp]
\center
\includegraphics[width=0.95\linewidth]{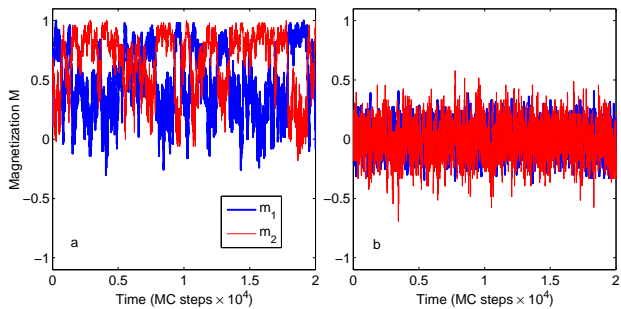} 
\caption{\footnotesize Typical time-evolution of the magnetizations
per spin of the two modules, $m_1$ and $m_2$, in (a) chimera ordered
state and (b) disordered state,
shown for MC simulations with $N =100$ at (a) $k_B T/J = 5$ and (b) 
$k_B T/J = 8$, respectively.}
\label{mag_time}
\end{figure}
\begin{figure}[tbp]
\center
\includegraphics[width=0.95\linewidth]{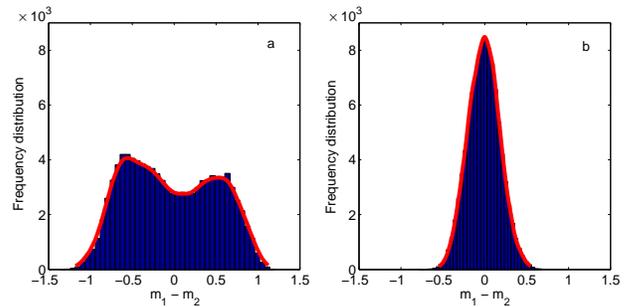} 
\caption{\footnotesize Frequency distribution of the quantity
$m_1-m_2$ shown when the spin system exhibits (a) chimera order and 
(b) disorder. The kernel smoothened distribution function is represented 
by a thick curve. Results shown for MC simulations with $N =100$ at (a)
$k_B T/J = 5$ and (b) 
$k_B T/J = 8$, respectively.}
\label{o_distrib}
\end{figure}

The other quantity is an 
order parameter $p=|m_1-m_2|$ that is used to identify a chimera state
in a spin system with two modules that have magnetizations $m_1$ and
$m_2$, respectively. 
To numerically estimate the order parameter $p$, we are confronted
with a few potential complications.
First, as the magnetizations of the two modules
are stochastically fluctuating (especially the one which is
disordered, i.e., having lower magnetization), we
can use their average values. However, this gives rise to an
additional problem as the modules can frequently exchange their
order/disorder status, especially when the modules are of the same size
[Fig.~\ref{mag_time}~(a)]. 
Thus, a method is required to measure $p$ that is unaffected
by the modules switching their magnetizations, while 
being able to determine if the time-averaged magnetizations of the two
modules are different (the signature of chimera order).
For instance, a simple time average of the absolute value of
differences between $m_1$ and $m_2$ will not give correct results
as one obtains a finite value (because of stochastic fluctuations) 
even when the time averaged magnetizations are same for both modules 
[Fig.~\ref{mag_time}~(b)]. Therefore, to estimate $p$ we have used the
following algorithm. First, the frequency distribution of $m_1-m_2$ is
computed. In the absence of chimera, the distribution is
unimodal [Fig.~\ref{o_distrib}(a)] which is approximated as a Gaussian
function, while for chimera ordering the
distribution is bimodal [Fig.~\ref{o_distrib}(b)] which is
approximated as a superposition of two Gaussian functions.
For a unimodal distribution, the value of $m_1-m_2$ corresponding to
the peak is used to calculate the order  parameter $p$. For bimodal
distributions, a 
weighted average of the values corresponding to the peaks is used. For
example, 
if one of the peaks occurs at $p_1$ with value $h_1$ while the other
is at $p_2$ with value $h_2$, the order parameter
is calculated as $p=(h_1|p_1|+h_2|p_2|)/(h_1+h_2)$. 
In order to make the automated detection of the peak locations accurate, the
frequency distributions need to be smoothed of all fluctuations in
the frequencies so that the local maxima at the peaks can be uniquely
determined. For this purpose we have used a kernel smoothing
technique~\cite{bowman} that gives a single peak location for unimodal
distributions and locations for two peaks in the case of bimodal distributions.

\section{Results}
\label{sec:results}
Starting from a structurally balanced network comprising two modules,
we first introduce field and thermal fluctuations so as to drive the
system into chimera order - i.e., one of the modules becomes ordered
while the other is disordered. Once this is achieved we allow adaptation
dynamics of the interactions to take place.
As the evolution of the interactions are related to the degree of 
fluctuations in the spin states, these would occur mostly in the
disordered module where the spins are subject to the competing forces
of negative interactions with the spins belonging to the ordered
module and the influence exerted by the uniform external
magnetic field which tries to align the spins in parallel with those
of the ordered module.
Thus, at any instant, a few spins in the disordered module will get
aligned with the spins in the other module and consequently may
develop positive interactions with the latter as a result of link
adaptation. This means that they will now no longer be part of the
disordered module but become part of the ordered module. As a result,
the size of the modules would change over time - the ordered module
increasing at the expense of the disordered one.
This has to be taken into account when measuring system properties,
such as, magnetization per spin of the two modules.
In order to have a consistent definition of module size that would be
valid even when the system is no longer in structural balance, we
heuristically measure it as follows.
At each update of the $J_{ij}$s, we randomly choose a spin from the
ordered module and consider all spins that have positive interactions
with it to belong to the ordered module, with the remaining spins
comprising the disordered module. Following this process, we can
follow the time-evolution of modular membership of individual spins, 
as well as, that of the module sizes. However, this measure becomes
less useful as the system becomes increasingly frustrated.

\begin{figure}[tbp]
\centering
\includegraphics[width=0.95\linewidth]{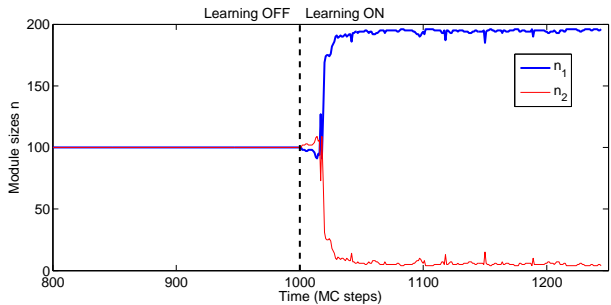}
\caption{\footnotesize Time-evolution of sizes of the modules
resulting from link adaptation dynamics that is introduced at time $t
= 1000$ MC steps (indicated by broken line). The ordered module
expands in size ($n_1$) while the disordered one shrinks.
Learning rate $\epsilon$ = 0.1}
\label{hebb_module_size}
\end{figure}

\begin{figure}[tbp]
\center
\includegraphics[width=0.95\linewidth]{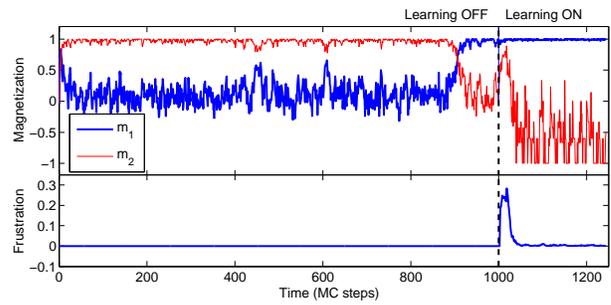}
\caption{\footnotesize (top) Time-evolution of magnetizations $m_1$ and
$m_2$ of the two modules in the network (having sizes $n_1$ and $n_2$,
respectively) before and after link adaptation dynamics is introduced
at time $t
= 1000$ MC steps (indicated by broken line).
In absence of adaptation, magnetizations show the characteristic
signature for chimera order. Once adaptation dynamics is operational,
the ordered module with magnetization $m_1$ is seen to increase in
size.
(bottom) In absence of learning, the network exhibits no frustration
as it is structurally balanced. Following the introduction of
adaptation dynamics, there is a transient increase in frustration,
before the system once again achieves balance, but with modules of
asymmetric sizes.}
\label{hebb_m1m2}
\end{figure}

\begin{figure}[tbp]
\center
\includegraphics[width=0.95\linewidth]{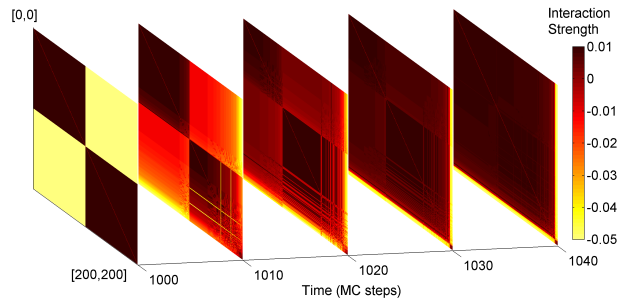}
\caption{\footnotesize Snapshots of the time-evolution of the
interaction matrix {\bf J} after adaptation dynamics is initiated at
$t = 1000$ MC steps. 
Note that the time interval shown here corresponds to the same period
over which the
frustration in the system initially rises and then decreases again as
the system once more reaches a balanced state, as indicated in 
Fig.~\ref{hebb_m1m2}.}
\label{jsnaps}
\end{figure}

Fig.~\ref{hebb_module_size} illustrates the evolution of module sizes
with time following the introduction of link adaptation dynamics in
the system. We can now measure the magnetizations per spin, $m_1$ and
$m_2$, of the two
modules by taking into account the modular membership of each spin and
the module sizes. Fig.~\ref{hebb_m1m2} shows
these order parameters, as well as, the frustration in the network, as
a function of time before and after the link adaptation dynamics is
introduced. 
As expected, we observe that following the start of link adaptation
dynamics, the ordered module begins to grow in size, which is also
observed in the time-evolution of the interaction matrix (shown as
snapshots at regular intervals in Fig.~\ref{jsnaps}).

\begin{figure}[tbp]
\center
\includegraphics[width=0.95\linewidth]{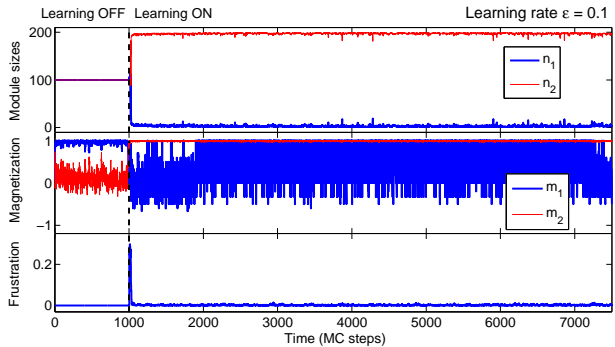}
\caption{\footnotesize 
Time-evolution of module sizes (top panel), magnetization (middle
panel)
and frustration (bottom panel) for an initially chimera ordered system
in the presence of an external field
before and after link adaptation dynamics with rate
$\epsilon=0.1$ is introduced (the time at which the
link adaptation is initiated is indicated by the broken line). 
Following a brief transient rise in the frustration, the system again
attains a balanced state although with asymmetric module sizes.
}
\label{eps1e-1}
\end{figure}

\begin{figure}[tbp]
\center


\includegraphics[width=0.95\linewidth]{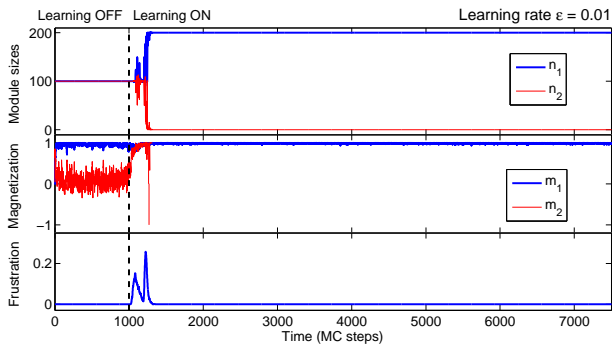}
\caption{\footnotesize Time-evolution of module sizes (top panel),
magnetization (middle
panel)
and frustration (bottom panel) for an initially chimera ordered system
in the presence of an external field before and after link adaptation
dynamics with rate $\epsilon=0.01$ is introduced (the time at which
the link adaptation is initiated is indicated by the broken line).
The system eventually becomes completely ordered with all interactions
becoming positive (resulting in merging of the two modules into a
single one).}
\label{eps1e-2}
\end{figure}

\begin{figure}[tbp]
\center
\includegraphics[width=0.95\linewidth]{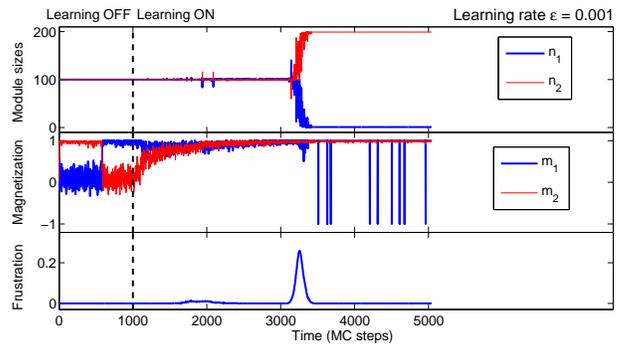}
\caption{\footnotesize Time-evolution of module sizes (top panel),
magnetization (middle panel) and frustration (bottom panel) for an
initially chimera ordered system in the presence of an external field
before and after link adaptation dynamics with rate $\epsilon=10^{3}$
is introduced (the time at which the link adaptation is initiated is
indicated by the broken line). Initially, the two modules both become
ordered. However, eventually the two modules merge into a single one
as all interactions become positive. The time of merging coincides
with the transient rise in frustration.
}
\label{eps1e-3}
\end{figure}

\begin{figure}[tbp]
\center
\includegraphics[width=0.95\linewidth]{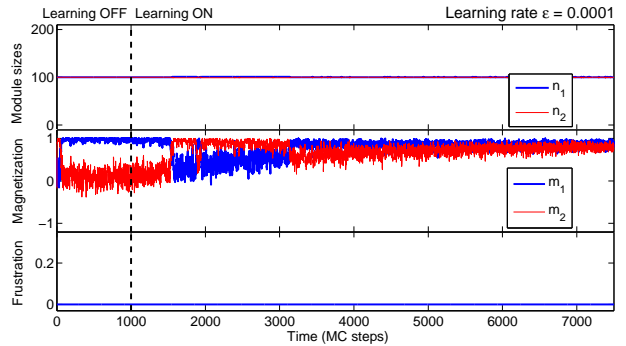}
\caption{\footnotesize Time-evolution of module sizes (top panel),
magnetization (middle panel) and frustration (bottom panel) for an 
initially chimera ordered system in the presence of an external field
before and after link adaptation dynamics with rate $\epsilon=10^{4}$
is introduced (the time at which the link adaptation is initiated is 
indicated by the broken line). We observe that while one of the
modules remains ordered, the initially disordered modules is gradually
also tending towards complete order. Eventually one expects the two
modules to merge into a single one.
}
\label{eps1e-4}
\end{figure}

The time-evolution of the initial chimera-ordered network subjected to link
adaptation dynamics with various learning rates ranging from $\epsilon
= 0.1$ to $10^{-4}$ are indicated in
Figs.~\ref{eps1e-1}-\ref{eps1e-4}.
For fast learning rates (e.g., $\epsilon = 0.1$), the system immediately
converges to a
balanced state characterized by a large ordered module (comprising
about $95 \%$ of all the elements) and a small disordered one. By
observing the magnetization time-series  of the two modules in
Fig.~\ref{eps1e-1} it is easy to infer that the system still exhibits
chimera order. We note that a rise in frustration is seen for a very
brief duration during the initial period immediately following
initiation of the link adaptation dynamics.
For a slower learning rate, viz. $\epsilon=0.01$, 
the process of expansion in size of the ordered module is different
(Fig.~\ref{eps1e-2}). We notice that soon after the beginning of link
adaptation, the chimera ordering is lost as the disordered module
becomes ordered. However, the modules still retain their different
identities as they are defined based on the sign of interactions
of the spins within them (which should be positive) and that with
spins in the other module (which should be negative).
For even slower learning rates such as
$\epsilon=10^{-3}$ or $10^{-4}$ (Figs.~\ref{eps1e-3}-\ref{eps1e-4})
the duration over which the two modules exist while both being ordered is seen
to increase. During this period the modular membership of the spins do
not change significantly (apart from small fluctuations). Chimera
ordering is lost as the interaction strengths for spins in the disordered module
becomes weaker as a result of frequent switchings of their
orientations. As a result the effect of the external field becomes
dominant, which causes the spins in the disordered module to align
with it. As spins in both modules are now aligned parallel to each
other,
adaptation of their interactions with time would eventually make all
interactions positive, thereby merging the two modules into one.
The time at which this happens is indicated by a transient rise in the
frustration (Figs.~\ref{eps1e-2} and \ref{eps1e-3}), occurring much
later for slower rates of adaptation.

The different scenarios we observe at fast and slow learning rates
suggest that there is a competition between two processes: 
if the modular membership of the spins can change rapidly following
initiation of link adaptation so that the system adopts a new
structurally balanced configuration, the interaction strengths of the
spins in the disordered module do not decrease significantly - thereby
allowing the two modules to coexist with the chimera ordering intact
(although the modules now have very different sizes).
Note that the modular membership of the spins will change rapidly
as they continually change their interaction strengths because of the
fast adaptation rate. However, the average number of spins in each
module will remain fairly steady.
If the learning rate is slow, 
the interaction strengths will not have
time to change sufficiently rapidly to allow the system to reach a new
structurally balanced state following the initiation of link
adaptation dynamics. As a result the field dominates over the weakened
interactions of the spins in the disordered module, thereby
eliminating chimera order and eventually causing the two modules to
merge.

\begin{figure}[tbp]
\center
\includegraphics[width=0.95\linewidth]{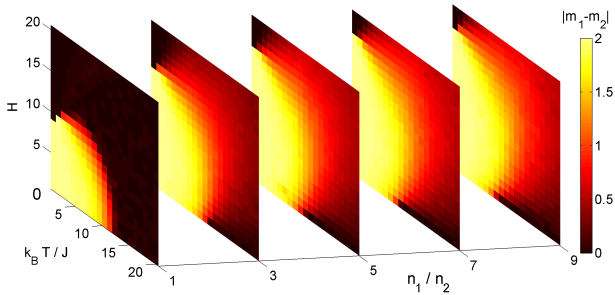}
\caption{\footnotesize The region of chimera order in the ($H, T$) parameter
space shown for different ratios of the sizes of the two modules.
Chimera is indicated in terms of high values of the order parameter
$|m_1-m_2|$ (represented by different colors).}
\label{ratio}
\end{figure}

The existence of chimera order even in situations where the two
modules have very different sizes is a novel observation as earlier it
had only been observed in a system with symmetric module
sizes~\cite{Singh2011}. In order to
see how the region in the field-temperature parameter space where we
observe chimera ordering varies with asymmetry in the module sizes, we
show in Fig.~\ref{ratio} the dependence of the order parameter
$|m_1-m_2|$ on the strength of the external field ($H$), the scaled
temperature ($k_BT/J$) and the ratio of sizes of the two modules
($n_1/n_2$) in the absence of link adaptation dynamics. We observe
that the region in
parameter space where we find chimera order (higher values of
$|m_1-m_2|$) increases significantly as we go towards higher module
size ratios (i.e., $n_1/n_2$) especially for lower values of
temperature.

\begin{figure}[tbp]
\center
\includegraphics[width=0.95\linewidth]{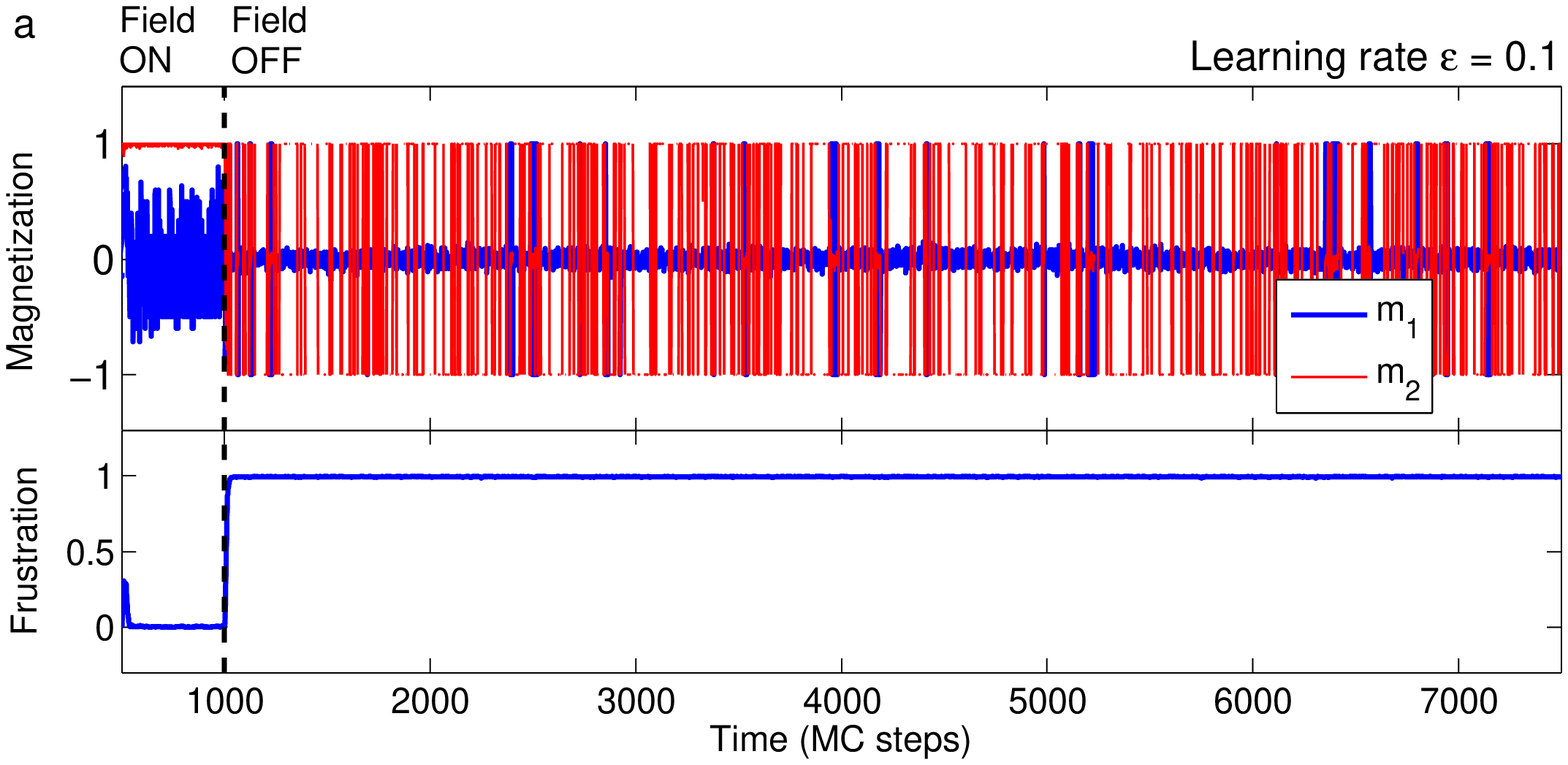}
\includegraphics[width=0.95\linewidth]{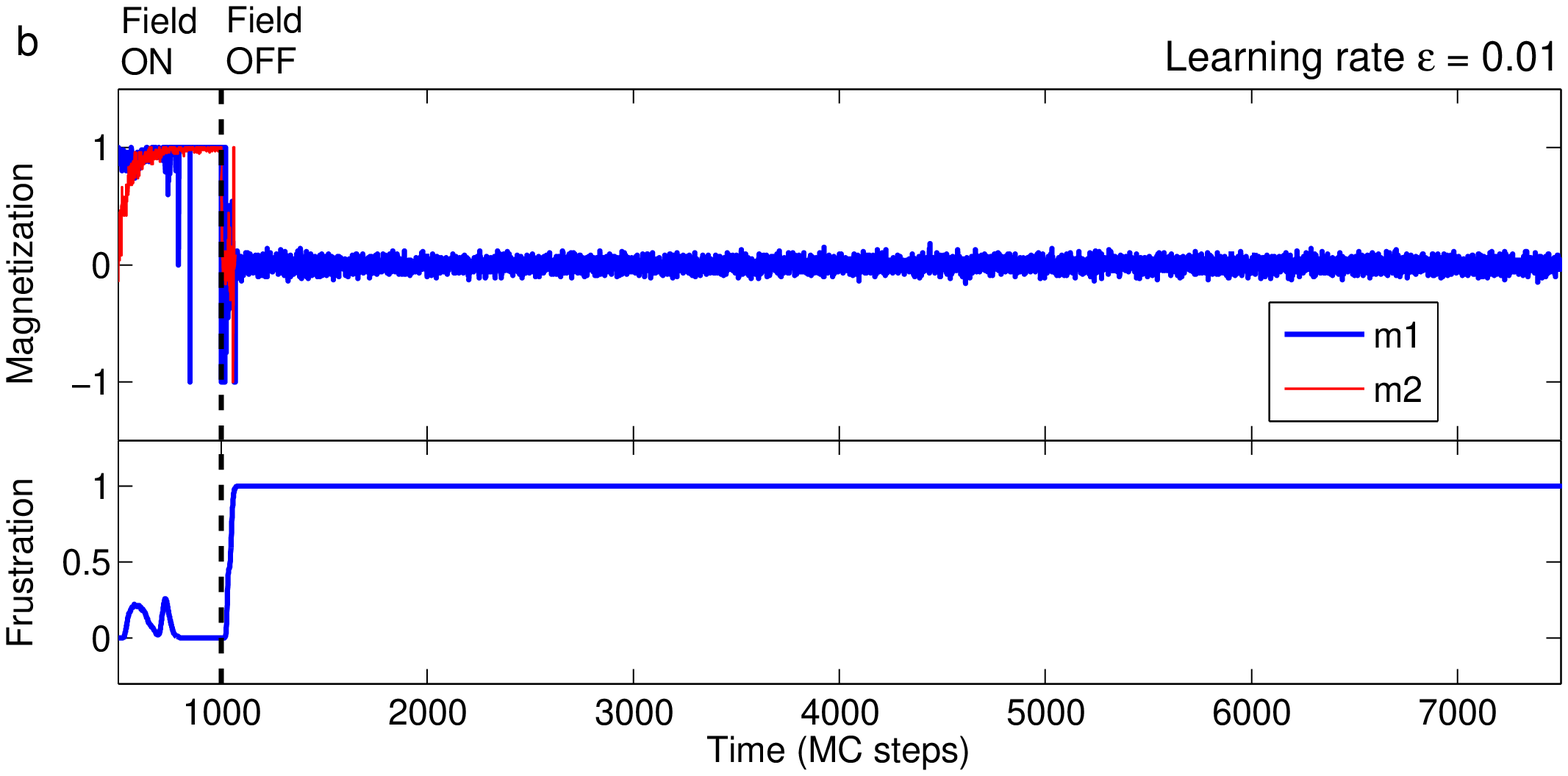}
\caption{\footnotesize (a) Time-evolution of magnetization (top panel)
and frustration (bottom panel) for an initially chimera ordered system
with link adaptation rate $\epsilon =0.1$,
before and after an external field is withdrawn (the time at which the
field is switched off is indicated by the broken line). We observe
that the system becomes completely frustrated on withdrawal of the
field. (b) The corresponding time-evolution shown for link adaptation
rate $\epsilon=0.01$.}
\label{off_eps1e-1-2}
\end{figure}

We now consider the situation when the external field is withdrawn
while allowing the link adaptation dynamics to continue.
A difference is expected with the results shown in the earlier
work~\cite{Singh2014} where the effect of link adaptation on the 
evolution of a system subject to
thermal fluctuations was investigated, as the learning rule is
different on account of the asymmetry in the strengths allowed for
negative and positive interactions. As the adaptation dynamics used
here allows for much stronger negative interactions (as compared to
positive), we would expect the system to become more frustrated as
negative interactions are much more likely to contribute frustrated
triads. This is indeed what is observed for slow learning rates, e.g.,
$\epsilon=0.1$ and $\epsilon=0.01$ (Fig.~\ref{off_eps1e-1-2}). 
Note that in a frustrated system, one cannot define modules in a
meaningful way, and thus the distinction between modules indicated in the
time-series for magnetization is an artifact of the measurement
method. For both of the slow learning rates we observe that the system
converges to a fully frustrated state corresponding to the maximum
value of the measure of frustration in the system ($=1$).

\begin{figure}[tbp]
\center
\includegraphics[width=0.95\linewidth]{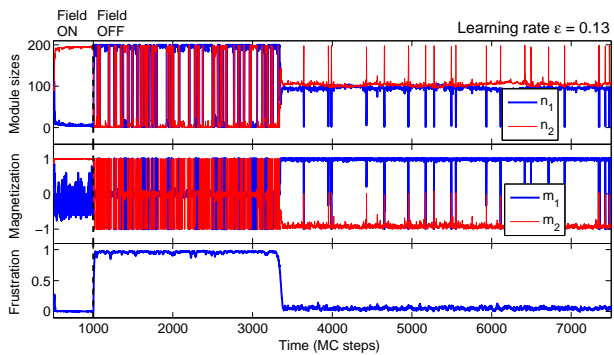}
\caption{\footnotesize 
Time-evolution of module sizes (top panel), magnetization (middle panel)
and frustration (bottom panel) for an initially chimera ordered system
with link adaptation rate $\epsilon =0.13$,
before and after an external field is withdrawn (the time at which the
field is switched off is indicated by the broken line). We observe
that the system initially becomes completely frustrated on withdrawal of the
field. However after about $2000$ MC steps the system suddenly becomes
balanced.}
\label{off_eps1.3}
\end{figure}

\begin{figure}[tbp]
\center
\includegraphics[width=0.95\linewidth]{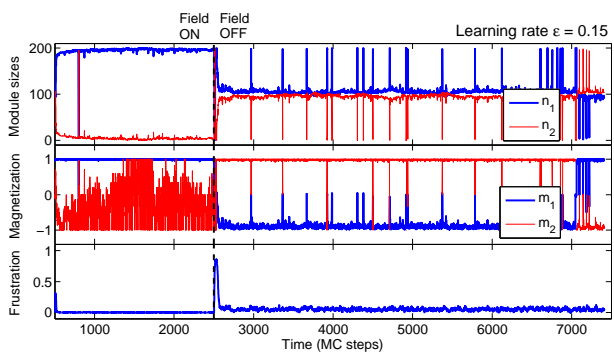}
\caption{\footnotesize 
Time-evolution of module sizes (top panel), magnetization (middle
panel)
and frustration (bottom panel) for an initially chimera ordered system
with link adaptation rate $\epsilon =0.15$,
before and after an external field is withdrawn (the time at which the
field is switched off is indicated by the broken line). We observe
that the system quickly becomes balanced after a brief transient rise
in frustration following the withdrawal of the field.
}
\label{off_eps1.5}
\end{figure}

For faster link adaptation rates, however, we observe unexpected behavior in
the system. For a learning rate $\epsilon = 0.13$ [Fig.~\ref{off_eps1.3}], which is only
slightly higher than $\epsilon = 0.1$ [shown in
Fig.~\ref{off_eps1e-1-2}~(a)], the system initially becomes fully
frustrated after the external field is withdrawn. 
Surprisingly, after a period of $\sim 2000$ MC steps during
which the system remains frustrated, 
it suddenly converges to a balanced state. This intervening period
between the field being switched off and the system spontaneously
splitting into
two modules
having positive interactions among all elements within them (and
correspondingly, negative interactions between elements belonging to
different modules) becomes even shorter with increasing learning rates
(e.g., see Fig.~\ref{off_eps1.5} for $\epsilon = 0.15$).
Thus, we can identify a critical value of around $0.13$ for the link adaptation 
rate $\epsilon$, below which the system remains frustrated and above
which the system converges to a balanced state, on withdrawal of the
external field. Indeed our simulations show that at $\epsilon = 0.13$
there is a wide variation in the time required for the system to
converge to balance after the field is switched off.

In the previous study of structural balance~\cite{Singh2014}, it had
been observed that the time required to converge to a structurally
balanced state exhibits a bimodal distribution for a range of
temperatures. We observe similar features in our model also - for
instance, for a link adaptation rate of $\epsilon = 0.13$. 
For slightly slower learning rates (e.g., $\epsilon = 0.12$), the
convergence time increases significantly. In our simulations, many
realizations did not converge to the balanced state within the time of
observation ($10^4$ MC steps). For higher learning rates, the system
rapidly converges to balance. However, unlike the convergence to
balance seen in the previous study~\cite{Singh2014}, the modules in
the structurally balanced state that our model system reaches are of
almost equal size (see Figs.~\ref{off_eps1.3} and \ref{off_eps1.5})
independent of the initial state of the network
(including the initial distribution of interactions which was seen to
affect the nature of the balanced state that the system converges to
in the earlier work).
This may appear surprising as starting from a balanced state, one
expects that the network will remain in that state as the interactions
adapt so as to make the corresponding energy minimum even deeper.
However, what we observe is that, on withdrawing the field the
balanced state characterized by asymmetric module sizes is
destabilized and the energy minimum moves in the configuration space
eventually reaching the region corresponding to balanced state with
modules having equal size. This difference between our results and
that of the previous study suggests that the adaptation rule used here
(which is biased in favor of negative interactions) is responsible for
an intriguing meta-dynamics of the energy landscape itself.

We have also observed how the structure
of the energy landscape underlying our system
evolves as the interactions change through the adaptation rule. 
In the model studied earlier~\cite{Singh2014}, the frustration of the
initially disordered system is around 0.5. On initiating link
adaptation, this value decreases as the system tends towards balance.
When frustration reaches a value around 0.49, we observe that the
energy landscape becomes such that starting from any spin
configuration one can reach the same minimum energy spin arrangement.
This suggests that a small decrease in the frustration can accompany a
a very large change in the basin of attraction of an energy minimum
(which now encompasses a significant fraction of the 
configuration space).
We observe similar behavior in our model system where the adaptation
occurs in the presence and subsequent absence of an external field.
For example, for an adaptation rate of $\epsilon=0.13$, when the field
is withdrawn the system initially becomes completely frustrated with
the value of frustration measured as around 0.97. The corresponding
energy landscape has a very large number of minima, each having very
small basins of attraction. As the system evolves towards a balanced
state, we note that even when the frustration decreases by a very
small amount, e.g., to 0.93, the energy landscape transforms to one
having an extremely deep energy minimum with a very large basin of
attraction. We believe that this radical transformation of the energy
landscape of the system at the initial stages of approach to balance
points to features of landscape evolution that deserve further study.

\section{Conclusions}
\begin{figure}[tbp]
\center
\includegraphics[width=0.95\linewidth]{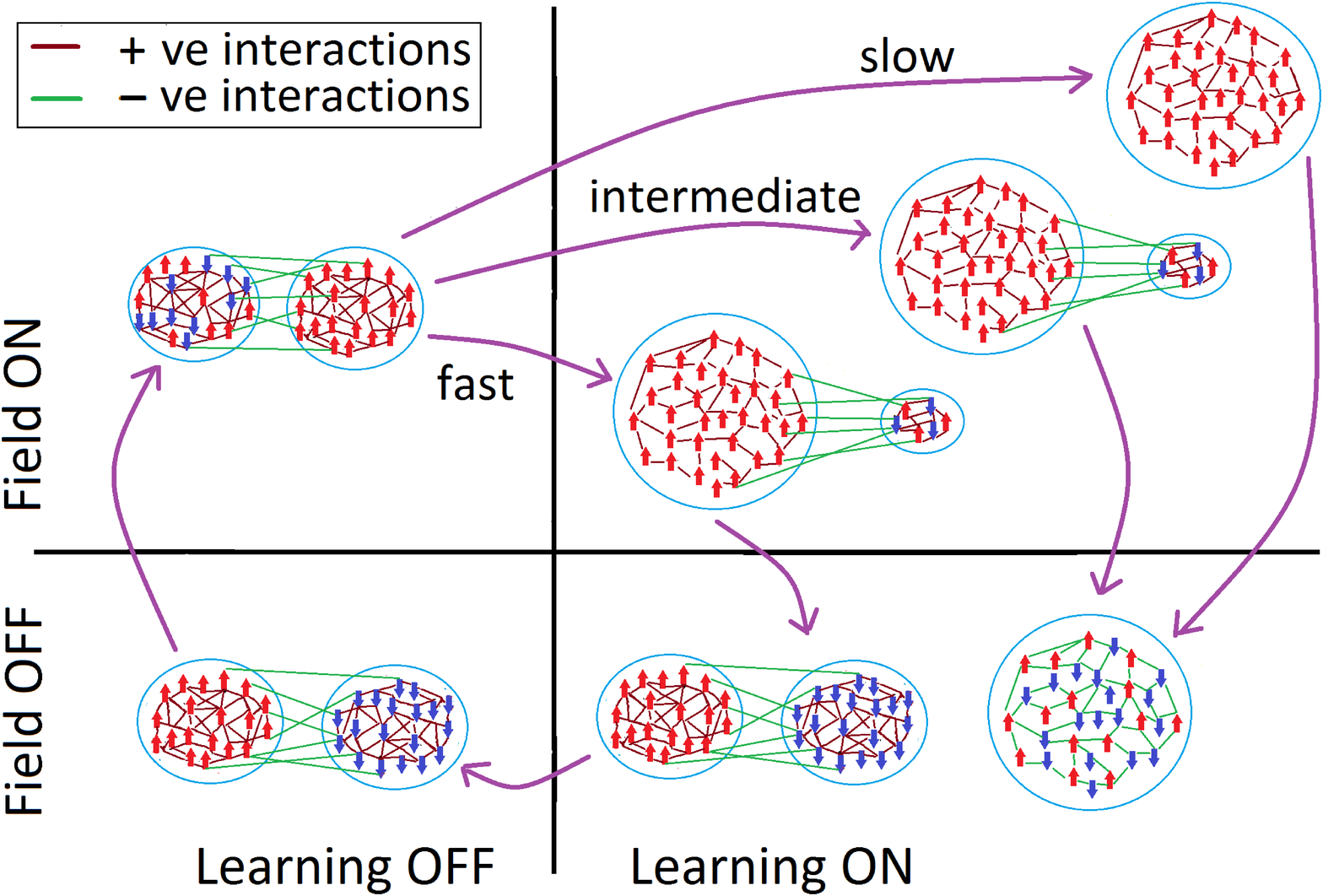}
\caption{\footnotesize Schematic representation of our key results.
Starting from the structurally balanced network (with the different
communities showing opposite ordering) at the bottom left
quadrant, we obtain chimera state on introducing an external field of
requisite strength (top left quadrant). Allowing the interactions to
adapt with different learning rates result in the network maintaining
balance but with different outcomes for the ordering. Fast
and intermediate learning rates retain chimera order, although the
disordered region is much diminished, while slow learning leads to
complete order (top right quadrant). On withdrawing the field, the
system subject to fast learning rate again converges to a
structurally balanced network with oppositely ordered
communities of approximately equal
size, similar to the original network. Intermediate and slow learning, however,
results in a fully frustrated network with complete disorder (bottom
right quadrant).
}
\label{schematic}
\end{figure}

In this paper we have explored the behavior of a spin system that is
subjected to an external field, while the interactions adapt in
response to the system dynamics.
The results obtained through our simulations that is described above 
is summarized in Fig.~\ref{schematic}.
From an earlier study, we had known that introducing link adaptation
inspired by the Hebbian principle in a frustrated system can result in
it converging to a structurally balanced state - corresponding to
evolution of the 
underlying energy landscape - that depends on the
initial state of the system. Applying this adaptation dynamics to a
system exhibiting chimera order (as done in here) allows us to explore
how the dynamics of the energy landscape is affected by an external
field, as well as, asymmetry in the adaptation rule regarding 
negative and positive interactions.

We observe that when system is adapting
through faster learning rates, the chimera order is maintained
throughout its evolution, with the ordered module increasing in size
at the expense of the disordered module.
For slower learning rates,
first the interactions of the elements belonging to the disordered
module become weak which results in
loss of chimera order. This is followed by the system converging to a fully
ordered state where almost all the interactions are positive. The sizes
of the two modules defining the system where one observes chimera
order can be varied to
see how the ratio of module sizes can affect the subsequent evolution of the
system. Indeed, we see that the region in the field-temperature
parameter space over which chimera order is seen increases as
the ratio increases from 1.
When the external field is withdrawn, the system returns to a 
structurally balanced state with modules of similar sizes if the
adaptation rate is high. However, for slower learning rates, the
system becomes completely frustrated. 

We have also tried to explore how the
structure of the energy landscape of the system
evolves during learning. Starting from a chimera ordered state that is
subjected to adaptation rate and external field, we observe changes in
the landscape after the field is switched off. Initially the system
becomes completely frustrated. However, after an interval (which
decreases with increasing learning rate) the system shows a sudden
large increase in the basin of attraction of an energy minimum,
although the frustration of the system has decreased only negligibly.
This surprisingly radical change in the landscape resulting from relatively
small degree of change in the interaction structure of the network is
a question that needs to be explored further.

It is of interest to consider the implications of the results reported
here for adaptation in biological systems. As connections in the brain
evolve according to long-term potentiation that embodies the Hebbian
principle that has inspired the link adaptation dynamics used in our
study, it is natural to expect that the brain may exhibit at least
some of the features observed here. Indeed, as mentioned earlier, the
experimental observation of two dynamically anti-correlated
subnetworks in the resting human brain~\cite{Fox2005} strongly suggests 
that in the
absence of strong external stimulation the underlying network is
structurally balanced.
On the other hand, when exposed to stimuli, correlated brain activity
is indeed observed - although not encompassing the entire network.
This is, to some extent, reminiscent of the chimera ordered state that
is seen in our model system. Thus, the brain may be seen as
corresponding to a system subject to relatively fast adaptation rate.
However, the Hebbian principle can apply to a much broader class of
biological systems, e.g.,
gene regulation networks where the co-expression of genes
has been suggested to result in their co-regulation over
evolutionary time-scales~\cite{Fernando2009}. As the adaptation in
this case is provided by natural selection, which is orders of
magnitude slower than the learning process in the brain mentioned
earlier, it is probably not unreasonable to conclude that this can be seen as
one corresponding to our model system subject to slow rate of
adaptation. It is easy to see that the frustrated system with a large
number of energy minima can be considered to be analogous to the
cellular differentiation process that allows convergence to any one of
a large number of possible cell fates dependent upon initial
conditions. Indeed this analogy has been used earlier by Kauffman to
motivate Boolean network models for explaining differentiation during
biological development~\cite{Kauffman1991}. Extending this analogy, one
may wonder whether the system has a state corresponding to the ordered
state seen on exposing it to an external stimulus. As this state is
unique and will be attained by the system independent of all initial
conditions, it is tempting to suggest that induced pluripotency in
cells exposed to chemical stimuli~\cite{Takahashi2006} 
may be the biological analogue.
Therefore, it may be of interest to study the phenomena reported here
in biologically realistic models of networks adapting at different
time scales.

\section*{acknowledgements}
We thank Rajeev Singh, Shakti N. Menon and R. Janaki for helpful discussions.
This work was supported in part by IMSc Complex Systems Project (XII
Plan) funded
by the Department of Atomic Energy, Government of India.
The numerical work was carried out in machines of the IMSc High-Performance
Computing facility, including ``Satpura'' which is
partially funded by DST (Grant No. SR/NM/NS-44/2009).

\bibliographystyle{amsplain}

\end{document}